\documentclass{phbauth}        
\usepackage{graphicx}

\begin{document}

\begin{frontmatter}

\title{Quantum phase transitions in antiferromagnets and superfluids}

\author{Subir Sachdev\thanksref{thank1}} and
\author{Matthias Vojta\thanksref{thank2}}

\address{Department of Physics, Yale University, P.O. Box 208120,
New Haven CT 06520-8120, USA}

\thanks[thank1]{E-mail: subir.sachdev@yale.edu}
\thanks[thank2]{E-mail: matthias.vojta@yale.edu}

\begin{abstract}
We present a general introduction to the non-zero temperature dynamic and
transport properties
of low-dimensional systems near a quantum phase transition.
Basic results are reviewed in the context of experiments
on the
spin-ladder compounds, insulating two-dimensional antiferromagnets,
and double-layer quantum
Hall systems. Recent large $N$ computations on an extended $t$-$J$
model (cond-mat/9906104)
motivate
a global scenario of the quantum phases and transitions in the
high temperature superconductors, and connections are made to
numerous experiments.
\end{abstract}


\end{frontmatter}

\section{Introduction}

The last decade has seen many experimental studies of the
spin dynamics of complex transition metal oxides. Many fascinating new phenomena
have been discovered (including high temperature
superconductivity), and our understanding of strongly correlated
electronic systems has been greatly enhanced. Section~\ref{ladders}
of this paper will
describe three broad classes of behavior that have been observed
in the spin dynamics: many, but not all, of these oxides fall into
one of these three classes. We will present this discussion in the
context of a description of the zero temperature, quantum phase
transition found in a simple, toy model of a two dimensional
antiferromagnet of $S=1/2$ Heisenberg spins. Recent theoretical
work on the different non-zero temperature regimes of spin
relaxation and transport in the vicinity of the quantum phase
transition will be described, and correlated with experimental
observations.

Related ideas apply also to non-zero temperature
{\em charge\/} transport in
two-dimensional systems in the vicinity of a zero temperature
superfluid-insulator transition; these will be be briefly
noted.

Section~\ref{dope} of this paper will consider recent theoretical work
on the interplay between superconductivity, spin-density-wave and charge-density-wave
order in a model of doped two-dimensional antiferromagnets. We
will argue that the results suggest a natural scenario under which
the transition in the spin sector (from a state with long-range
magnetic order to a quantum paramagnet) falls in the same
universality class as the simple, undoped insulating model
considered in Section~\ref{ladders}; this scenario also
offers a natural explanation for numerous experiments on the cuprate
superconductors. We will
also briefly mention charge-ordering transitions in this doped
antiferromagnet, and their possible relationship to the anomalous
photoemission linewidths observed in a recent experiment.

We will close in Section~\ref{qhe} by discussing related magnetic
transitions in bilayer quantum Hall systems.

\section{Coupled ladder antiferromagnet}
\label{ladders}

We will begin our discussion by describing the basic physical
properties of a simple quasi two-dimensional model of $S=1/2$
Heisenberg spins. Our motivation is to introduce essential
concepts in the theory of quantum phase transitions~\cite{book,physworld}, and to
provide a crude picture of the spin fluctuations in the cuprate
superconductors---this picture should not be taken too literally
though, and a more precise correspondence will be made later in
Section~\ref{dope}. Further, there exist insulating transition
metal oxides~\cite{azuma} which are described by spin models closely related to
the one we consider here.

\begin{figure}[tb]
\begin{center}\leavevmode
\includegraphics[width=1.0\linewidth]{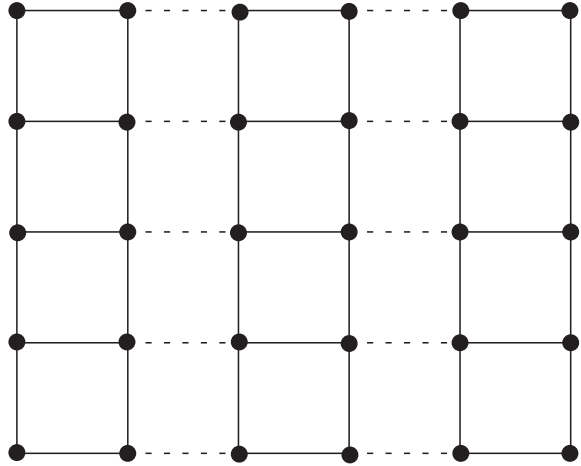}
\caption{
The coupled ladder antiferromagnet. Spins ($S=1/2$) are placed on the sites,
the $A$ links are shown as full lines, and the $B$ links as dashed
lines.
}\label{fig1}\end{center}\end{figure}

We consider the Hamiltonian
\begin{equation}
H = J \sum_{i,j \in A} {\bf S}_i \cdot {\bf S}_j + \lambda J
\sum_{i,j \in B} {\bf S}_i \cdot {\bf S}_j
\label{ham}
\end{equation}
where ${\bf S}_i$ are spin-1/2 operators on the sites of the
coupled-ladder lattice shown in Fig~\ref{fig1}, with the $A$ links
forming
decoupled two-leg ladders while the $B$ links couple the ladders as
shown.
The ground state of $H$ depends only on the dimensionless
coupling $\lambda$, and we will describe the low temperature
properties as a function of $\lambda$.

For simplicity, we will restrict our attention in this section
to
the regime $J>0$, $0 \leq \lambda \leq 1$.

Let us first consider the case where $\lambda$ is close to 1.
Exactly at $\lambda=1$, $H$ is identical to the square lattice
Heisenberg antiferromagnet, and this is known to have long-range,
magnetic N\'{e}el order in its ground state {\em i.e.} the
spin-rotation symmetry is broken and the spins have
a non-zero, staggered, expectation value in the ground state
with
\begin{equation}
\langle {\bf S}_i \rangle = \eta_i N_0 {\bf n},
\end{equation}
where ${\bf n}$ is some fixed unit vector in spin space,
and $\eta_i$ is $\pm 1$ on the two sublattices.
This long-range order is expected to be preserved for a finite
range of $\lambda$ close to 1. The low-lying excitations above the
ground state consist of slow spatial deformations in the orientation
${\bf n}$ in the form of spin waves. There are {\em two} polarizations
of spin waves at each wavevector $k = (k_x, k_y)$ (measured from the
antiferromagnetic ordering wavevector), and they have excitation energy
$\varepsilon_k = \hbar (c_x^2 k_x^2 + c_y^2 k_y^2)^{1/2}$, with $c_x, c_y$
the spin-wave velocities in the two spatial directions.

\begin{figure}[tb]
\begin{center}\leavevmode
\includegraphics[width=0.8\linewidth]{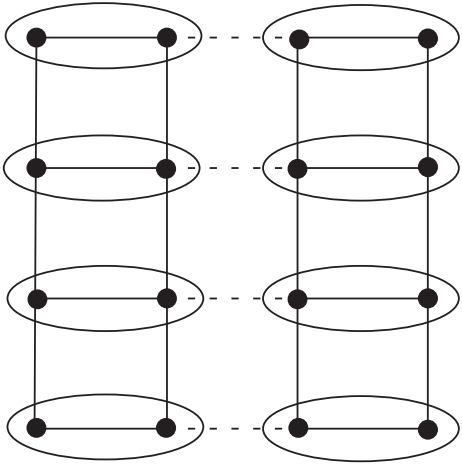}
\caption{
Schematic of the quantum paramagnet ground state for small
$\lambda$. The ovals represent singlet valence bond pairs.
}\label{fig2}\end{center}\end{figure}

Let us turn now to the vicinity of $\lambda = 0$. Exactly at
$\lambda=0$, $H$ is the Hamiltonian of a set of decoupled spin
ladders. Such spin ladders are known to have a paramagnetic ground
state, with spin rotation symmetry preserved, and an energy gap to
all excitations~\cite{dagotto}. A caricature of the ground state
is sketched in Fig~\ref{fig2}: spins on opposite rungs of the ladder
pair in valence bond singlets in a manner which preserves all
lattice symmetries.
Excitations are now formed by breaking a
valence bond, which leads to a {\em three}-fold degenerate state
with total spin $S=1$; this broken bond can hop from site-to-site,
leading to a triplet quasiparticle excitation.
For $\lambda$ small, but not exactly 0, we expect that the ground
state will remain a gapped paramagnet, and the excited
quasiparticle will now move in two dimensions.
We parameterize its
energy at small wavevectors by
\begin{equation}
\varepsilon_k = \Delta + \frac{\hbar^2 (c_x^2 k_x^2 + c_y^2 k_y^2)}{2
\Delta},
\label{epart}
\end{equation}
where $\Delta$ is the energy gap, and $c_x$, $c_y$ are velocities.

\begin{figure}[tb]
\begin{center}\leavevmode
\includegraphics[width=1.0\linewidth]{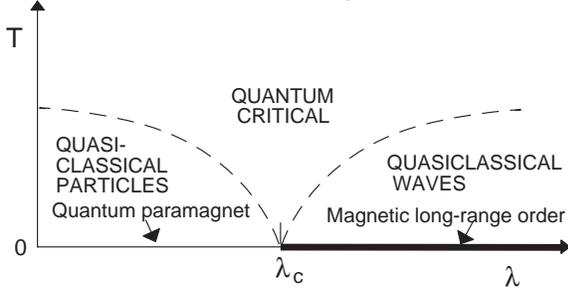}
\caption{
Phase diagram of $H$ for $T>0$ and $0 \leq \lambda \leq 1$.
The dashed lines represent
crossovers.
}\label{fig3}\end{center}\end{figure}

The very distinct symmetry signatures of the ground states and excitations
between $\lambda \approx 1$ and $\lambda \approx 0$
make it clear that the two limits cannot be
continuously connected. It is known~\cite{katoh,twor} that there
is an intermediate second-order phase transition at $\lambda = \lambda_c \approx
0.3$. Both $\Delta$ and $N_0$ vanish continuously as $\lambda_c$
is approached from either side. The following subsections will
consider the distinct physics
at very low $T$ for $\lambda > \lambda_c$ and $\lambda < \lambda_c$
respectively (Fig~\ref{fig3}).
We will find that in both cases a quasiclassical
description of the long time dynamics is possible, although the
effective classical models are very different. This will be
followed by a discussion of the higher temperature `quantum
critical' region, where quantum and classical effects are not as
easy to disentangle.

\subsection{Quasiclassical waves}
\label{qw}

We consider $\lambda > \lambda_c$, and $T$ very small. This regime
was considered in detail in Refs~\cite{CHN}.

There are two key observations. \\
({\em i}) The non-linear interactions
between the thermally excited spin waves lead to loss of long-range
N\'{e}el order at any $T>0$. The order parameter correlations
decay on the scale of the correlation length, $\xi$, which obeys
\begin{equation}
\xi \sim \exp ( 2 \pi \rho_s / k_B T),
\end{equation}
where $\rho_s$ is the geometric mean of the spin stiffnesses towards
deformations of the ground state in the two directions;
a thermodynamic argument shows that $\hbar^2 c_x c_y = \rho_s /\chi_{u
\perp}$, where $\chi_{u \perp}$ is the $T=0$ susceptibility to a uniform
magnetic field oriented in a direction orthogonal to the N\'{e}el
order (both $\rho_s$ and $\chi_{u \perp}$ vanish as
$\lambda \searrow \lambda_c$, but $c_x$, $c_y$ remain
finite).\\
({\em ii})
The most important spin waves occur at a wavevector $k \sim
\xi^{-1}$, and at this scale, their thermal occupation number is
large:
\begin{equation}
\frac{1}{e^{\varepsilon_k /k_B T} - 1} \gg 1.
\label{bose}
\end{equation}
Consequently, a classical description of these spin wave modes
with large occupation numbers is possible. \\
A continuum Hamiltonian theory of
non-linearly interacting waves can be written down in terms of the
instantaneous orientation of the N\'{e}el order, ${\bf n} (x, t)$,
and the conserved uniform magnetization density ${\bf L} (x,t)$.
The $T>0$ response functions of the antiferromagnet are obtained
after integrating over a classical thermal ensemble of initial
conditions for ${\bf n}$ and ${\bf L}$;
the initial conditions for ${\bf L} (x)$ are assumed to be given by a
Gaussian with variance controlled by $\chi_{u \perp}$. Note that
the allowed values of ${\bf L}$ are continuous and there is no
signature of quantization of spin---this is a consequence of the
large occupation number of the elementary quantum modes in
(\ref{bose}).

A key quantity characterizing the quasiclassical wave dynamics is
the phase coherence time $\tau_{\varphi}$, loosely defined as the time over which
memory of the local orientation of the N\'{e}el order parameter is
lost. This can be estimated by the equation of motion for ${\bf
n}$:
\begin{equation}
\hbar \frac{\partial {\bf n}}{\partial t} = \frac{1}{\chi_{u \perp}}
{\bf L} \times {\bf n},
\label{e1}
\end{equation}
which states that ${\bf n}$ precesses about the local
magnetization. We can now estimate that
\begin{equation}
\langle | {\bf L} | \rangle \sim \sqrt{ \langle {\bf L}^2 \rangle } \sim \sqrt{ k_B T
\chi_{u \perp} /\xi^2},
\label{e2}
\end{equation}
where the last estimate follows from assuming that ${\bf L}$ fluctuations
have energy $k_B T$ over a scale $\xi$, and the classical
fluctuation-dissipation theorem.
From (\ref{e1}) and (\ref{e2}) we may estimate that time over
which phase memory is lost is
\begin{equation}
\tau_{\varphi} \sim \hbar \sqrt{\chi_{u \perp}/k_B T} \, \xi.
\label{tp1}
\end{equation}
This relaxation time controls linewidths in dynamic neutron scattering
experiments, and also the relaxation rates in NMR; for the latter
we have $1/T_1 \sim \tau_{\varphi}$, leading to exponentially
large values of $1/T_1$ as $T \rightarrow 0$~\cite{CO}.

\subsection{Quasiclassical particles}
\label{qp}

Now we consider $\lambda < \lambda_c$, and $T$ small.

As argued in Ref~\cite{SD}, a quasiclassical model for the long
time dynamics is again possible, but it now involves particles
carrying integer quanta of spin, rather than waves with a
continuous magnetization. The particles are of course those with
excitation energy (\ref{epart}), and at finite $T$ they appear
with a density
\begin{equation}
\rho = 3 \int \frac{d^2 k}{4 \pi^2} e^{-\varepsilon_k /k_B T} \sim
e^{-\Delta/k_B T}.
\end{equation}
The particles have a kinetic energy of order $k_B T$, and
therefore a thermal de-Broglie wavelength of order $1/\sqrt{T}$.
So as $T \rightarrow 0$, the particle spacing becomes
exponentially large, and eventually exceeds their de-Broglie
wavelength, suggesting that they can be treated classically.

The two-dimensional spin dynamics of these quasi-classical
particles can be described in a $1/N$ expansion
by a Boltzmann equation~\cite{book}. However, for the case where
the motion of the particles is quasi one-dimensional {\em i.e.} for $\lambda
=0$ or with $\lambda$ small and the observation time shorter than
the inter-ladder hopping time, some exact results are possible.
We write the magnetization density associated with the thermally
activated particles as a sum over particles carrying discrete
spin quanta
\begin{equation}
L_z (x, t) = \sum_k m_k \delta (x - x_k (t)),
\end{equation}
where $x_k (t)$ is the trajectory of particle $k$ which carries
azimuthal spin $m_k = 1, 0, -1$ (contrast this with the continuous
spin distribution assumed for ${\bf L}$ in Section~\ref{qw}).
The classical Liouville equation for the
evolution of the ensemble of $x_k (t)$ can be solved
exactly~\cite{SD,DS}, and implications for the spin correlations
of the underlying antiferromagnet can then be computed, and are
described below.

All results in this paragraph are for the case of quasi
one-dimensional motion.
The interparticle interactions lead to a broadening of the
quasi-particle pole in the dynamic structure factor (measured in
neutron scattering). This broadening occurs on a time scale,
$\tau_{\varphi}$,
of the order of the time between particle collisions, and this is
\begin{equation}
\tau_{\varphi} \sim \frac{\hbar}{k_B T} e^{\Delta/k_B T}.
\label{tp2}
\end{equation}
The collisions also lead to diffusive transport of spin, and a
spin diffusion constant $D_s$ given by
\begin{equation}
D_s = \frac{\hbar c_y^2}{3 \Delta} e^{\Delta/k_B T},
\end{equation}
where we have assumed that the one-dimensional motion is along the
$y$ direction. This diffusive transport can be very clearly
detected in NMR experiments~\cite{takigawa}. In an external field,
$H$, spin diffusion leads to a $1/\sqrt{H}$ dependence in the
$1/T_1$ relaxation rate. The absolute value of $1/T_1$ can also be
computed:
\begin{equation}
\frac{1}{T_1} = \frac{\Gamma \Delta e^{-3 \Delta/2 k_B T}}{c_y^2}
\sqrt{ \frac{3 k_B T}{\pi H}},
\label{qp1}
\end{equation}
where $\Gamma$ is a hyperfine coupling. The same NMR experiment
measuring $1/T_1$ can also detect the uniform susceptibility by
the Knight shift, and the present quasiclassical particle model
predicts that to be
\begin{equation}
\chi_u = \frac{e^{-\Delta/k_B T}}{\hbar c_y} \sqrt{\frac{2 \Delta}{\pi
k_B T}},
\label{qp2}
\end{equation}
per unit length of the spin ladder.
A striking property of (\ref{qp1},\ref{qp2}) is worth emphasizing:
note that the $T \rightarrow 0$ activation gap of $1/T_1$ is 3/2
times the activation gap in $\chi_u$. Indeed, this is very close
to the experimental trends observed in a large number of quasi
one-dimensional spin gap systems, as summarized in
Ref~\cite{yasu}. More detailed quantitative comparisons of the theoretical
predictions have been performed~\cite{DS}, including the ballistic
to diffusive crossover in the particle motion at time scales of
order $\tau_{\varphi}$, and the agreement with experiments is
quite satisfactory.

\subsection{Quantum critical dynamics}
\label{qc}

We now raise the temperature into the quantum critical region,
shown in Fig~\ref{fig3}.

For $\lambda < \lambda_c$, this region is reached when $k_B T \sim
\Delta$. From the arguments above, we see that, for
$k_B T \sim \Delta$, the quasiparticle
spacing is of order their de-Broglie wavelength, and so a
quasiclassical particle picture cannot be generally applicable.
Similarly, for $\lambda > \lambda_c$, quantum criticality is
reached when $k_B T \sim \rho_s$, and then the occupation number of
the important spin-wave modes is of order unity, and a direct
quasiclassical wave theory becomes invalid. Thermal and quantum
fluctuations are therefore equally important in the quantum
critical region, and a simple, intuitive, classical picture of the
dynamics does not exist. Rather, progress in our understanding has
come from a combination of scaling arguments, and expansions
based on $\epsilon = 3-d$ ($d$ is the spatial dimensionality)
and $1/N$ ($N$ is the number of order parameter components; $N=3$
for the antiferromagnets being considered here).
Further, in some restricted regimes of $d$, $N$, or frequency,
$\omega$, either a particle or a wave-like picture becomes more
appropriate, and further simplifications of the dynamical theory
then become possible~\cite{book,rd,oleg}.

The key property of the quantum-critical regime is that a suitably
defined phase coherence time obeys~\cite{SY,CSY}
\begin{equation}
\tau_{\phi} \sim \frac{\hbar}{k_B T},
\label{tp3}
\end{equation}
where the missing proportionality constant is a universal number.
This result can be obtained by taking the limiting boundary values
of (\ref{tp1},\ref{tp2}) as $T$ is raised into the
quantum-critical regime. More generally, (\ref{tp3}) follows from
general scaling arguments, and the fact that $k_B T$ is the most
important low energy scale in this regime. This energy scale
also controls the value of other observables: the uniform magnetic
susceptibility (per unit area) now obeys~\cite{CS,CSY}
\begin{equation}
\chi_u = \Omega \frac{k_B T}{\hbar^2 c_x c_y}
\end{equation}
where $\Omega$ is a universal number, and the NMR relaxation rate
is given by~\cite{CS,CSY}
\begin{equation}
\frac{1}{T_1} \sim T^{\eta},
\end{equation}
where $\eta$ is an exponent close to 0.

The quantum-critical transport properties are also of some
interest. We will express the results in terms of spin
conductivity, $\sigma$. Unfortunately, it has so far not been
possible to measure $\sigma$ in two-dimensional antiferromagnets,
but we hope measurements will appear in the future. However, the properties
of {\em charge} transport near a superfluid-insulator transition
of Cooper pairs are very similar, obey the same scaling forms,
and are more easily accessible in experiments. The conductivity
obeys~\cite{cond,fgg}
\begin{equation}
\sigma (\omega, T) = \frac{Q^2}{h} \Sigma \left( \frac{\hbar
\omega}{k_B T} \right)
\end{equation}
where $Q= g \mu_B$ for spin transport and $Q=2e$ for charge
transport, and $\Sigma$ is a universal scaling function describing
the crossover in the dynamical conductivity at a frequency scale
of order $\tau_{\varphi}^{-1}$. Both $\Sigma(0)$ and $\Sigma (\infty)$
are expected to be finite universal constants, with distinct
values. The value $\Sigma (0)$ describes the incoherent d.c.
transport in which pre-existing, thermally excited quasiparticles
undergo hydrodynamic drift in the applied field; in contrast $\Sigma (\infty)$
describes energy absorbed by pairs of excitations created by
an oscillating external field.

\section{Doped antiferromagnets}
\label{dope}
This section will highlight aspects of some recent
results~\cite{vojta}
on the
ground state phase diagram of doped square lattice
antiferromagnets. We will focus on the nature of some quantum
transitions in the model, their relationship to the simple
model we discussed in Section~\ref{ladders}, and to
experiments on the cuprate superconductors~\cite{emery}.

\begin{figure}[tb]
\begin{center}\leavevmode
\includegraphics[width=1.1\linewidth]{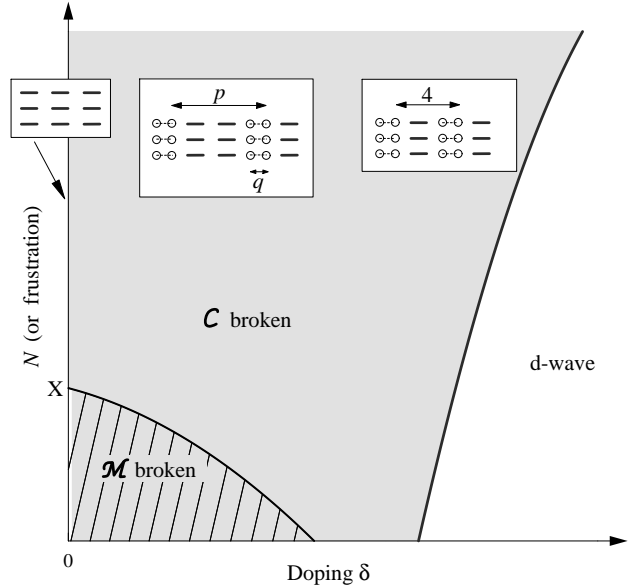}
\caption{
Ground state phase diagram of a doped antiferromagnet, adapted from
Ref~\protect\cite{vojta}. The magnetic ${\mathcal M}$ symmetry is broken in the
hatched region, while ${\mathcal C}$ symmetry is broken
in the shaded region; there are numerous additional phase transitions at which
the detailed nature of the ${\mathcal M}$ or ${\mathcal C}$ symmetry breaking changes - these
are not shown. For $\delta=0$, ${\mathcal M}$ symmetry is broken
only below the critical point $X$, while ${\mathcal C}$ symmetry is broken only
above $X$. The superconducting ${\mathcal S}$ symmetry
is broken for all $\delta > 0$ at large $N$;
for smaller $N$, the ${\mathcal S}$ can be restored at small $\delta$
by additional ${\mathcal C}$ breaking along the vertical axis for the
states in the inset--this is not shown.
The nature of the ${\mathcal C}$ symmetry breaking at large $N$
is sketched: the thick and dashed lines indicate varying
values of the bond charge density, while the circles represent site hole density.
}\label{fig4}\end{center}\end{figure}

Ref.~\cite{vojta} considered the phase diagram (Fig~\ref{fig4}) of a model
of doped
antiferromagnets on the square lattice, as functions of the doping
$\delta$ and a second parameter, $N$, which can be loosely
interpreted as the number of components of each spin.
It is also believed that moderately frustrating non-nearest neighbor exchange
interactions will have an effect similar to increasing the value
of $N$, and so the vertical axis of Fig~\ref{fig4} can also be
considered to be proportional to the strength of such frustration.

The focus in the computation was on the fate of three distinct
symmetries of the Hamiltonian in the ground state. There are
numerous competing instabilities associated with breaking one or
more of these symmetries, and this competition leads to intricate
possibilities in the phase diagram. The three symmetries are:
\begin{itemize}
\item ${\mathcal S}$, the electromagnetic $U(1)$ gauge symmetry. This
is broken in any superconducting state.
\item ${\mathcal M}$, the $SU(2)$ spin symmetry. This is broken in
magnetically ordered states which break spin rotation invariance.
\item ${\mathcal C}$, the symmetry of the space group of the square
lattice. We will only consider this broken if an observable
invariant under ${\mathcal S}$ and ${\mathcal M}$ does not respect square
lattice symmetries. In particular, ${\mathcal C}$ will be broken if
the charge density on every bond and site is not identical.
Thus the two-sublattice N\'{e}el state at $\delta = 0$ breaks only
${\mathcal M}$ while an incommensurate, collinear spin density wave
breaks both ${\mathcal M}$ and ${\mathcal C}$.
\end{itemize}
An important characteristic of the phases we describe below is that
knowledge of the structure of ${\mathcal S}/{\mathcal M}/{\mathcal C}$
breaking gives us essentially complete information on the nature
of the ground state and its low energy excitations. In other
words, knowing the symmetry breaking, we can perform a canonical
electron Hartree-Fock/BCS small fluctuation analysis, and identify
all the important low energy modes. Additional gapless modes with
singular interactions and temperature dependencies appear only at
quantum critical points between the phases, and these are
therefore of central interest.

We first discuss the phases at $\delta =0$. For small $N$, the
ground state is the two-sublattice N\'{e}el state. For larger $N$,
there is a spin-Peierls state with a $2 \times 1$ unit cell which
breaks ${\mathcal C}$ only (this state has a modulation in the bond charge
density). There is evidence~\cite{kotov,book} the transition between these states
is second order, and that the ${\mathcal M}$ and ${\mathcal C}$ order
parameters vanish continuously as the critical point is approached
from opposite sides. It is interesting that a similar phenomenon
happens in frustrated spin chains in $d=1$, where Ising/N\'{e}el
and spin-Peierls order vanish continuously at the same point and
do not co-exist~\cite{hald}. It appears, therefore, that there is
a `duality' between ${\mathcal M}$ and ${\mathcal C}$ order at $\delta =
0$.

Let us now consider non-zero doping  $\delta > 0$. One of the key points
made in Ref~\cite{vojta} is that it is simpler to first consider
doping the spin-Peierls state found at larger $N$. This has the
advantage of ensuring that ${\mathcal M}$ symmetry remains unbroken,
and we only have to consider the competition between two
symmetries, ${\mathcal C}$ and ${\mathcal S}$.
Such a procedure will be sensible if the ${\mathcal C}$-ordering
instabilities of the cuprate superconductors were, in a sense, more
fundamental than the ${\mathcal M}$-ordering. There is experimental evidence
this is the case (the ${\mathcal C}$-ordering scattering peaks appear
at a higher temperature than the ${\mathcal M}$-ordering)
and this supports our approach.

Our results are shown in Fig~\ref{fig4}. At low doping one
invariably finds one-dimensional striped structures in which the
holes are concentrated into regions which are an even number of
sites ($q$) wide. The distance between stripes ($p$) is inversely
proportional to the doping, $\delta$, for small $\delta$,
as is found in the cuprate
superconductors~\cite{emery}.
The even width is preferred because it promotes the
pairing of spins into singlet bonds. This pairing leads to strong
superconductivity along the longitudinal stripe direction; there
is a weaker Josephson tunneling coupling in the transverse
direction, and this eventually leads to two-dimensional superconductivity.
Thus the predominant feature of the small $\delta >0$, and large
$N$,
portion of the phase diagram is the co-existence of ${\mathcal C}$
and ${\mathcal S}$ breaking. This co-existence is instrinsically a
two-dimensional feature. In one dimension, quasi-long-range
charge and superconductivity orders are dual to each other, and
do not coexist in a `Luther-Emery' liquid. However, in two
dimensions, the additional directional freedom allows coexistence:
the charge-density wave ordering is in the $x$ direction, while
the superconductivity is predominantly along the longitudinal $y$
direction. At smaller $N$ and small $\delta$, it is possible that
additional charge density wave ordering will appear also in the
longitudinal direction---we expect the resulting state to then be
an insulator. Conversely, at large $\delta$, the ${\mathcal C}$
breaking disappears completely (Fig~\ref{fig4}),
and we obtain an ordinary $d$-wave
superconductor.

The computation above also allows one to follow the fermionic
excitation spectrum in the ${\mathcal C}$ and ${\mathcal S}$
broken phases. In the large doping $d$-wave superconductor, there
are the familiar gapless fermionic excitations at special singular
points along the $(1,\pm 1)$ directions. When ${\mathcal C}$ order
initially appears (and assuming the absence of time-reversal symmetry
breaking), these points move away from the
special directions (additional gapless points also appear at
image points separated by the new reduced reciprocal lattice
vectors). However, when the strength of the ${\mathcal C}$ order is
large enough, the gapless points disappear, and the fermion
spectrum is gapped over the entire Brillouin zone. Note that at no
value of $\delta$ is there a sign of a large Fermi
surface--instead we have at best gapless Fermi points. This will
be crucial in our discussion below of quantum phase transitions.

We now consider the situation at smaller values of $N$ and $\delta >
0$. The most important new physics is the breaking of ${\mathcal M}$
symmetry and the appearance of magnetic order. Unlike the
situation at $\delta =0$, we do not now expect a `duality' between
${\mathcal M}$ and ${\mathcal C}$ order, but expect them to
co-exist. This is as in $d=1$, where static impurities have been
shown to induce a staggered magnetization in ${\mathcal C}$ broken
states~\cite{fuku}. The magnetic order is expected in the form of
an incommensurate, collinear spin density wave, which breaks both ${\mathcal C}$
and ${\mathcal M}$ symmetries. We expect that the nature of the ${\mathcal C}$
symmetry breaking will not change very much at the $\delta > 0$
transition at which ${\mathcal M}$ symmetry is broken
(Fig~\ref{fig4}).

\subsection{Quantum phase transitions}
\label{qpt}

Let us begin at large $\delta$ where the ground state is a
$d$-wave superconductor. Upon decreasing $\delta$ we observe two
significant quantum phase transitions which we shall discuss further here: the
the initial onsets of ${\mathcal C}$ and ${\mathcal M}$ symmetry
breaking (Fig~\ref{fig4}).

Clearly, the spin/charge-density wave order is the appropriate
order parameter for these quantum transitions. In deriving the
proper quantum field theory for such an order parameter,
the key issue is whether there
are any additional low energy excitations which couple efficiently
to them. If we were considering the onset of
spin/charge-density wave order from a Fermi liquid~\cite{hertz,book},
then gapless fermionic excitations at Fermi surface points
separated by the ordering wavevector, ${\bf K}$, are key: they lead to a
damping of the order parameter modes and determine the
universality class of the transition. In the present situation, we
have no Fermi surface due to the ubiquity of superconductivity;
there are gapless fermionic excitations at isolated points in the
Brillouin zone, and so the central issue is whether any pair of
these points
are separated by a wavevector which equals ${\bf K}$ (or its
integer multiple) or not.

If the answer to the last question is no, then fluctuations of the
spin/charge density order do not scatter fermions between two
low lying states; they couple only to short-lived, virtual, fermion
particle-hole pair excitations. Such a coupling is quite innocuous
and serves only to renormalize the effective couplings in a field
theory for the charge/spin density wave order parameter in which
the fermionic degrees of freedom have been integrated out. In
particular, for the onset of ${\mathcal M}$ symmetry breaking, the
form of the effective action will be identical to that
for the quantum phase transition in the coupled ladder model
studied in Section~\ref{ladders}. In this case, most of the
results in the distinct dynamical regimes of Sections~\ref{qw},
\ref{qp}, and \ref{qc} can be applied essentially unchanged to the
present model. Related results will also hold for the onset
of ${\mathcal C}$ order.

In contrast, if ${\bf K}$ does equal the distance between two
gapless Fermi points, a somewhat different situation obtains. Now
a theory of order parameter modes coupled with the gapless fermion
excitations is necessary. A theory of this type has been studied
recently by Balents {\em et al}~\cite{balents}, who consider the
onset of ${\mathcal M}$ ordering in a $d$-wave superconductor. A
related model for ${\mathcal C}$ ordering was proposed in
Ref~\cite{vojta}. These theories are expected to possess a quantum
scale-invariant critical point with the exponent $\eta$
not necessarily close to 0. However, general features of the
finite temperature crossovers are expected to be similar to those
in Section~\ref{ladders}.

\subsection{Implications for cuprate superconductors}

First, let us consider the ${\mathcal M}$ ordering transition.
There are a number of experiments in ${\rm La}_{2-x} {\rm Sr}_x {\rm Cu O}_4$
which are consistent with the interpretation that such a transition,
with collinear incommensurate magnetic order, happens at around $x \approx
0.11$.
\begin{itemize}
\item
The early NMR measurements of Imai {\em et al} examined the
$T$ dependence of $1/T_1$ for a variety of values of $x$; their
measurements neatly fall into the three classes of behavior
discussed in Sections~\ref{qw},~\ref{qp}, and~\ref{qc}, with the
last being consistent with an $\eta \approx 0$. More recent NQR
measurements~\cite{hunt} are sensitive to charge order, and are
consistent with the topology of the phase diagram sketched in
Fig~\ref{fig4}.
\item
The neutron scattering measurements of Aeppli {\em et
al}~\cite{aeppli}, as well as earlier work~\cite{hayden,keimer},
indicate quantum critical scaling in the dynamic spin structure
factor with dynamic exponent $z \approx 1$ and $\eta \approx 0$.
\item
Recent NMR measurements of the transverse relaxation time,
$T_2$~\cite{fujiyama}, also indicate $z=1$ criticality.
\item
Numerous neutron scattering experiments~\cite{rossat1,mook,tony3,bourges}
have observed a sharp,
high energy `resonance peak' near the antiferromagnetic ordering
wavevector. Although a momentum-dependent dispersion of this peak
has not (yet) been observed, it is tempting to identify it with
the quasiclassical particles of Section~\ref{qp}.
\end{itemize}

We turn next to the onset of ${\mathcal C}$ order. There are no
direct signatures yet of critical fluctuations associated
with such a
transition~\cite{cdw}.
However, it has been proposed in Ref~\cite{vojta} that
the anomalous momentum linewidths observed in a recent
photoemission~\cite{photo}
experiment could be due to the scattering of the fermions from
critical ${\mathcal C}$ order fluctuations, as obtains in the theory discussed
in the latter part of Section~\ref{qpt}.

\section{Quantum Hall bilayers}
\label{qhe}

In closing, we briefly mention another experimental system in
which magnetic ordering transitions appear to have been observed;
in this case, a number of experimental knobs can rather easily
tune the parameters in the Hamiltonian, and the prospects from more
detailed experimental studies are good.

\begin{figure}[tb]
\begin{center}\leavevmode
\includegraphics[width=0.8\linewidth]{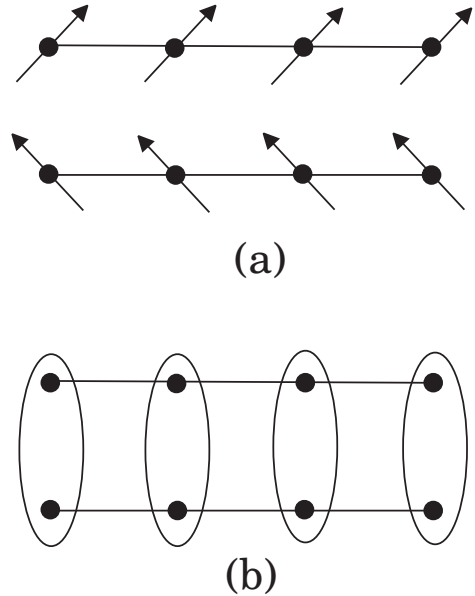}
\caption{
Schematic of the distinct ground states in the double layer
quantum Hall system at $\nu = 2$; the state $(a)$ is magnetically
ordered, while $(b)$ is a spin-singlet quantum paramagnet.
}\label{fig5}\end{center}\end{figure}

The transitions have been observed in a bilayer quantum Hall system at
filling fraction $\nu = 2$~\cite{pelle,sawada}. The electrons in
each layer occupy a fully filled Landau level at $\nu=1$, and so
it is not unreasonable in a first study to focus simply on the
spin degrees of freedom. Indeed, a reasonable caricature of the
physics can be obtained by imagining that the electrons are
rigidly fixed on equivalent lattices in each layer, and by describing
their interactions with an insulating spin model, as in (\ref{ham});
such a model has ferromagnetic exchange interaction within each
layer ({\em i.e.} the links in set $A$ in (\ref{ham}) couple nearest neighbor
sites within each layer and have $J < 0$), and antiferromagnetic exchange between
the layers (the links in set $B$ couple the two layers and have $\lambda J >
0$). The limiting ground states of such a Hamiltonian can now be
understood in manner similar to the discussion in
Section~\ref{ladders}, and are sketched in Fig~\ref{fig5}; for large
$|\lambda|$ the ground state is a quantum paramagnet in which the
spins in opposite layers pair to form singlets, while magnetically
ordered states are present at small $\lambda$.
A number of theoretical
studies~\cite{we1,DSZ,ZSD,troyerss,ezawa,demler1,demler2,macd,demler3}
of such spin models, and others which include the orbital motion in
the Landau levels, have appeared, and a quantitative confrontation
between theory and experiment should be possible in the future.

\begin{ack}
We thank Kedar Damle, Sankar Das Sarma, Robert Laughlin,
Jan Zaanen, and Shoucheng Zhang for useful discussions and
collaborations. This research was supported by US NSF Grant No DMR
96--23181 and by the DFG (VO 794/1-1).

\end{ack}


\end{document}